# Velocity selective bi-polarization spectroscopy for laser cooling of metastable Krypton atoms

Y. B. Kale*, V. B. Tiwari, S. Singh, S. R. Mishra and H. S. Rawat

*Laser Physics Applications Section,*
*Raja Ramanna Centre for Advanced Technology, Indore - 452013, India.*

*Corresponding author: yogeshwar@rrcat.gov.in

We report a velocity selective bi-polarization spectroscopy (VS-BPS) technique to generate a background-free, dispersion-like reference signal which is tunable over a wide range of frequency. In this technique, a pair of linearly polarized weak probe beams passing through a gas cell of metastable Krypton (Kr*) atoms, overlaps with a pair of counter-propagating circularly polarized strong pump beams derived from an independently tunable control laser. The polarization spectroscopy signals from the two probe beams, after subtraction, result in VS-BPS signal. The spectral shifting in VS-BPS signal can be achieved by tuning the frequency of the control laser. The dependence of the amplitude and slope of the VS-BPS signal on the RF power used for excitation of Kr atoms in the gas cell and on the power of pump beams has been studied. The frequency stability of a diode laser locked with VS-BPS signal has been found to be better than the frequency stability of the laser locked with a saturated absorption spectroscopy (SAS) signal. The VS-BPS signal is finally used for stabilization and tuning of the cooling laser frequency for a magneto-optical trap (MOT) for Kr* atoms.

## 1. INTRODUCTION

High resolution nonlinear spectroscopic techniques such as saturated absorption spectroscopy (SAS) and polarization spectroscopy (PS) [1] have applications in resolving the narrow sub-Doppler spectral lines [2-5], as well as in frequency stabilization of the lasers commonly used for laser cooling of atoms [6]. The SAS technique [7] measures the decrease in the absorption of a weak probe beam in the presence of a strong pump beam, whereas PS technique [8, 9] measures the change of polarization state of the probe beam induced by a polarized pump beam. The advantage with PS, as compared to SAS, is its much better sensitivity and signal-to-noise ratio, apart from the sharp dispersion-like spectral profile obtained without any laser frequency modulation and phase sensitive detection [10-13]. The PS signal, however, contains the background which is susceptible to the temperature and intensity fluctuations. The background level in the PS signal can be eliminated by using bi-polarization spectroscopy (BPS) [14] which is a difference based PS technique [15-17].

The spectroscopic techniques those can provide the narrow spectral feature in their signals include PS spectroscopy, Autler-Townes (AT) splitting [18], Electromagnetically Induced Transparency (EIT) [19] and generation of odd-derivatives of the signal using lock-in amplifier [20]. The signals obtained with these techniques are more suitable for frequency locking due to higher stability of the lock arising from the higher slope of the locking signal. But smaller frequency width of the signal, as obtianed in these techniques, also limits the tuning range of the lock point. For example, the frequency tuning range in PS technique is limited to a range equal to the power-broadened transition linewidth. Thus a tuning of laser frequency beyond this range has to be achieved through frequency shifting by acousto-optic modulators (AOM's) [21]. This may however result in significant reduction of the power in laser beams passing through the AOM's. Thus a frequency stabilization technique based on PS with large tuning range would be more suitable. To achieve this, a locking signal, proportional to detuning from an atomic transition has been generated earlier by taking the ratio of dispersion and absorption spectra [22]. This, however, consists of an involved scheme of four photodetectors and an analog electronic circuit which performs necessary subtractions and division in real time.

In this work, we present the velocity selective bi-polarization spectroscopy (VS-BPS) technique to generate a sharp, dispersion like spectroscopy signal, which provides a wide range of tunability of the frequency lock-point of an external cavity diode laser (ECDL) system. In a commonly used PS technique, the pump and probe laser beams are having same frequencies and are generated from a single laser system. We used here a different approach in which pump and probe laser beams are obtained from different laser systems operating at different frequencies. We observe that by changing the pump laser frequency, the PS signal of probe beam can be tuned over a wide range. Experiments using PS technique with independent lasers have been earlier demonstrated to achieve a polarized velocity selective spectroscopy [23, 24], optically induced polarization rotation [25] and two-photon polarization spectroscopy [26-29]. A. Hernández-Hernández et al. in their work reported in ref. [23], scanned the pump laser frequency while keeping the probe laser frequency locked to a resonant frequency. The detection of probe beam intensity in this scheme allows an unambiguous determination of the atomic lines by discriminating these from the cross-over peaks. In our method, a pair of linearly polarized weak probe beams passing through a gas cell of Kr* atoms, overlaps with a pair of counter-propagating circularly polarized strong pump beams derived from an independently tunable laser (referred hereafter as control laser). The polarization spectroscopy signals obtained from the two probe beams, on subtraction, generate the VS-BPS signal which can be frequency tuned by the control laser. The dependence of the amplitude and slope of the VS-BPS signal on RF power used for excitation of Kr atoms in the gas cell and on the power of pump beams from control laser is studied. The performance of the VS-BPS signal for the locking of the ECDL system has been studied and also compared with that of a standard SAS signal. Using our VS-BPS spectrometer, the detuning of cooling laser used in a Kr* magneto-optical trap (MOT) has been varied to study the dependence of the number and temperature of cold atoms in the MOT on cooling laser detuning.

This article is organized as follows: Section 2 discribes the basic principle of VS-BPS. In section 3, the experimental arrangement for the realization of the VS-BPS in Kr* atoms is discussed. Our experimental observations and results are discussed in section 4. In this section, the parametric dependence of the observed VS-BPS signal has been discussed with the RF power and the control laser power. In the second part of the section 4, frequency stabilization using the VS-BPS signal is compared with the conventional SAS method. Towards the end of the section 4, the paper shows application of VS-BPS technique to a Kr* MOT [30] in our laboratory. Finally, the conclusion is presented.

## 2. VELOCITY SELECTIVE BI-POLARIZATION SPECTROSCOPY (VS-BPS)

In polarization spectroscopy (PS), a dispersive signal is generated when a linearly polarized weak probe beam undergoes a phase retardation due to circularly polarized pump beam induced birefringence in the medium. The optical pumping due to circularly polarized strong pump beam results in unequal distribution of pupulation in different Zeeman sub-levels. This leads to unequal refraction and absorption for two oppositely circularly polarized components of electric field in the linearly polarized probe beam.

The electric field of a linearly polarized probe beam, at frequency ω and wave-vector k, propagating in an atomic medium along z-axis is given as,

$$\mathbf{E} = \mathbf{E_0} e^{i(\omega t - kz)}; \; \mathbf{E_0} = (E_0, 0, 0) \quad (1)$$

This linearly polarized probe beam can be decomposed into $\sigma^+$ and $\sigma^-$ circularly polarized components as,

$$\mathbf{E}^\pm = \frac{1}{2} E_0 (\hat{\mathbf{x}} \pm i\hat{\mathbf{y}}) e^{i(\omega t - k^\pm z)} \quad (2)$$

where, + and − signs correspond to $\sigma^+$ and $\sigma^-$ components of the probe beam, and, $\hat{\mathbf{x}}$ and $\hat{\mathbf{y}}$ are the unit vectors in x and y directions respectively. In the presence of $\sigma^+$ pump beam, the above two circularly polarized probe components propoagate in the atomic medim with difference in their absorption coefficients, $\Delta\alpha_+$ and difference in their refractive indices, $\Delta n_+$ given by [1],

$$\Delta\alpha_+ = (\alpha_+^+ - \alpha_+^-) = \frac{\Delta\alpha_0}{1+x^2}, \quad (3)$$

$$\Delta n_+ = (n_+^+ - n_+^-) = \frac{c}{\omega_0} \Delta\alpha_0 \frac{x}{1+x^2}. \quad (4)$$

with x is defined as,

$$x = \frac{\omega - \omega_0}{\Gamma/2} \quad (5)$$

where, $\omega_0$ is frequency of transition, $\Delta\alpha_0$ is the difference in absorption of $\sigma^+$ and $\sigma^-$ components of the probe beam at line center, and $\Gamma$ is the power broadened linewidth. The difference $\Delta\alpha_+$ describes the circular dichroism which makes the probe light elliptically polarized. The difference $\Delta n_+$ describes the gryotropic birefringence which rotates the axis of polarization of the probe beam. The refractive index and absorption of the windows of gas cell with thickness d is defined by a complex quantity ($n_w^{*\pm}$), given by,

$$n_w^{*\pm} \cdot (2d) = b_r^\pm - i\frac{c}{2\omega}\alpha_w^\pm. \quad (6)$$

where, $b_r^\pm$ and $\alpha_w^\pm$ represent the refractive and absorptive parts.

The PS signal (for single pump beam $\sigma^+$), which is the transmitted intensity of the probe beam propagated through an input polarizer, the pump induced birefringent atomic medium and finally through an output analyzer, can be expressed as [14],

$$S_{PS}(\omega) = \frac{1}{4} I_0 e^{-(\alpha_+ L + \alpha_w)} \left[ \xi + (e^{2\Delta_{i+}} - e^{-2\Delta_{i+}}) - 2\cos 2(\theta' + \Delta_{r+}) \right] \quad (7)$$

where, $\Delta_{i+} = \frac{1}{4}\Delta\alpha_+ L + \Delta\alpha_w$, $\Delta_{r+} = \frac{\omega}{2c}\Delta n_+ L$, $\theta' = \theta + \frac{\omega}{2c}\Delta b_r$, with $\Delta\alpha_w = \alpha_w^+ - \alpha_w^-$, L is length of atomic medium, $\theta$ is the angle of output analyzer axis with respect to its direction when it was crossed with input polarizer and $\xi$ (<10$^{-6}$) is the residual transmission when the polarizer and analyzer are completely crossed.

The equation corresponding to $S_{PS}(\omega)$, can also be derived for PS with a $\sigma^-$ circularly polarized pump beam. After subtraction of this from Eq. (7) we can obtain the expression for BPS. In case of using the $\sigma^-$ pump beam, the difference in absorption coefficient and the difference in refractive indices of two oppositely circularly polarized probe beam components becomes $\Delta\alpha_- = -\Delta\alpha_+$; $\Delta n_- = -\Delta n_+$ and $\alpha_- = \alpha_+ = \alpha_0$. This is due to the symmetry in the transition strengths of magnetic sub-level transitions from $+m_F$ to $-m_F$ for $\sigma^+$ and $\sigma^-$ polarized light [31]. Therefore, the dispersive signals generated by two probe beams are of opposite slope due to oppositely polarized pump beams ($\sigma^+$ and $\sigma^-$) corresponding to these probe beams. The subtraction of these opposite slope PS signals causes the dispersive signals to add and the background to cancel. Under typical experimental conditions, the changes $\Delta\alpha_+, \Delta n_+, \Delta\alpha_w$ and $\Delta b_r$ are very small and therefore the subtraction results in the background-free sharp dispersion-like BPS signal [14] given by,

$$S_{BPS}(\omega) = I_0 e^{-(\alpha_0 L + \alpha_w)} \left[ \frac{1}{4} \Delta\alpha_0 L \Delta\alpha_w \frac{1}{1+x^2} + \Delta\alpha_0 L \frac{x}{1+x^2} \sin 2\theta' \right] \quad (8)$$

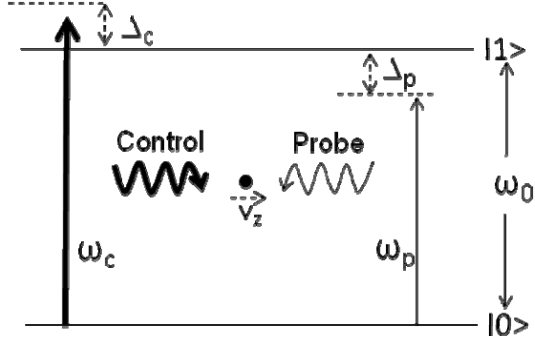

Fig. 1. Simple two level model for the understanding of VS-BPS in counter-propagating pump-probe geometry. The pump laser is control laser here.

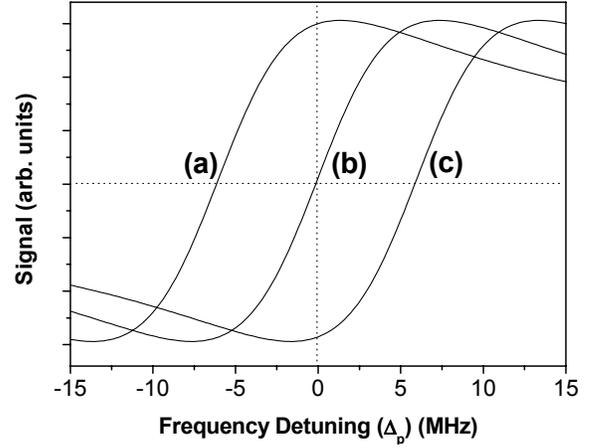

Fig. 2. Simulated variation in VS-BPS signal with $\Delta_p$ for different values of control laser detuning ($\Delta_c$). (a) $\Delta_c$ = 6 MHz ($v_z$ = 4.9 m/s), (b) $\Delta_c$ = 0 MHz ($v_z$ = 0 m/s), and (c) $\Delta_c$ = -6 MHz ($v_z$ = -4.9 m/s). Here, $\Delta\alpha_w$ = 0.08 and $\Gamma$ = 14.8 MHz.

The advantage of BPS over PS is that it allows for large $\theta$ (~ π/4) for generation of a large background-free dispersive signal.

In velocity selective bi-polarization spectroscopy (VS-BPS), the counter propagating pump and probe beams are considered at different frequencies. We consider a gaseous medium having two level atomic system (see Fig. 1) with finite Zeeman degeneracy in each of these level. The control laser pump beams are kept at at frequency $\omega_c$. The atoms with $v_z$ component of velocity **v** along the propagation direction of control laser beam (i.e. z-direction) can interact resonantly with this beam if,

$$\omega_c = \omega_0 + k v_z \quad (9)$$

These atoms with $v_z$ velocity component will be optically pumped by the control laser pump beam to generate the birefringence for probe beam at frequency $\omega_p$ given as,

$$\omega_p = \omega_0 - k v_z = 2\omega_0 - \omega_c \quad (10)$$

or

$$\Delta_p = -\Delta_c \quad (11)$$

where, $\Delta_p = \omega_p - \omega_0$ is detuning of the probe laser and $\Delta_c = \omega_c - \omega_0$ is detuning of the control laser. Thus when the control laser is locked to blue detuned from the atomic transition frequency, the probe laser will generate PS singal at red-detuned side from the transition frequency.

The generation of VS-BPS signal is based on the subtraction of the dispersion-like signals of weak probe beams obtained in a polarization spectroscopy setup in presence of velocity selective optical pumping (VSOP) [32] occurring due to control laser pump beams. Using straight forward algebra, the Eq. (8) can be transformed to obtain the VS-BPS signal for different probe and control (i.e. pump) laser frequencies as,

$$S_{VSBPS}(\omega_c, \omega_p) = I_0 e^{-(\alpha_0 L + \alpha_w)} \left[ \frac{1}{4} \Delta\alpha_0 L \Delta\alpha_w \frac{1}{1+y^2} + \Delta\alpha_0 L \frac{y}{1+y^2} \sin 2\theta' \right] \quad (12)$$

where, $y = 2(\omega_p + kv_z - \omega_0)/\Gamma = 2(\omega_p + \omega_c - 2\omega_0)/\Gamma$.

The Fig. 2 shows simulated variation in VS-BPS signal (Eq. (12)) with $\Delta_p$ for the different values of $\Delta_c$. We can note that different velocity class of atoms will be in resonance with control laser as its frequency is tuned around the transition frequency. This results in the smooth shifting of zero-crossing point in VS-BPS signal with change in control laser frequency. In Fig. 2, the VS-BPS signals obtained for different values of $\Delta_c$ correspond to different velocity class of atoms (resonant to the pump and probe beams simultaneously).

### 3. EXPERIMENT

The Kr atom has complete occupancy in its outermost ($4p^6$) shell. The relevent energy levels of Kr atom are shown in Fig. 3. We have used Radio Frequency (RF) excitation method (frequency ~ 30 MHz) to excite Kr to its first metastable state for the VS-BPS of Kr* in this work [33]. The excited metastable state $4p^55s[3/2]_2$ has lifetime ~ 40 seconds. The excitation from $4p^55s[3/2]_2$ to $4p^55p[5/2]_3$ can be realized using an ECDL of wavelength 811.5nm. The even isotopes (i.e. $^{78,80,82,84,86}$Kr) do not have hyperfine structure unlike the odd isotopes (i.e. $^{83,85}$Kr). The even-isotope $^{84}$Kr atom has the highest natural abundance (~57%).

The schematic of the experimental setup for realization of VS-BPS in Kr* atoms is shown in Fig. 4. The setup employs two ECDLs (TOPTICA, Germany) at wavelength 811.5 nm as control (ECDL (control)) and probe (ECDL (probe)) lasers respectively. A small part of laser power from both the ECDLs is utilized for the generation of SAS signals using Kr* gas cells filled at pressure of ~200 mTorr. The SAS signal generated from the setup SAS-R is used for the reference of the frequency. The SAS signal from the setup SAS-C is used for locking the control laser. The probe laser is locked using the VS-BPS signal generated from the setup shown in Fig. 4 after difference amplifier (DA). For comparision of VS-BPS technique with SAS for the locking of probe laser, the VS-BPS setup has been modified to generate SAS signal. For this, the control laser beam is blocked and the setup is modified such that pump and probe beams for SAS are available from the same probe laser. The major part of the probe laser power is used for the generation of cooling beams for the Kr* MOT. Here, the control and the probe laser beams ($1/e^2$ size ~ 0.5 mm) are in counter-propagating geometry as shown in Fig. 4. The probe beams (Pr1 and Pr2) generated by a thick glass plate (GP) are separated by a distance ~ 1cm. This separation between the probe beams is goverened by the thickness of the glass plate. A pair of orthogonally circularly polarized ($\sigma^+$ and $\sigma^-$) pump beams (Pu1 and Pu2) are generated by passing the output beam of the control laser through the polarizing beam splitters (PBS1 and PBS2) and λ/4 plate as shown in Fig. 4. These pump beams are made to counter-propagate to the direction of probe beams using mirror (M) and beam splitter (BS).

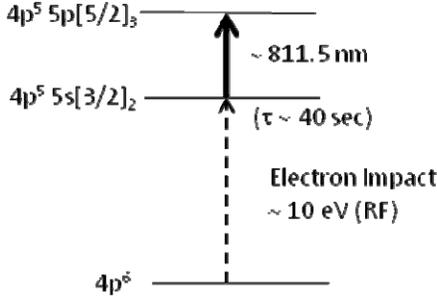

Fig. 3. Excitation scheme of Krypton atom.

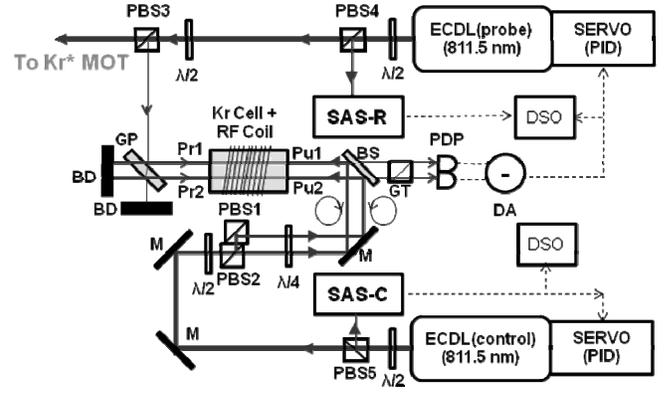

Fig. 4. The schematic of the Kr* bi-polarization spectroscopy setup which includes. ECDLs: external cavity diode lasers, BS: beam splitter, M: mirror, PBS: polarizing beam splitter, λ/2: half-wave plate, λ/4: quarter-wave plate, GP: glass plate, GT: Glan-Thompson prism, PDP: photodetector pair (in difference form), DSOs: digital storage oscilloscopes, DA: difference amplifier, BD: beam dump, Pu1 ($\sigma^+$), Pu2 ($\sigma^-$): Pump beams, Pr1, Pr2: Probe beams and SAS-R: saturated absorption spectroscopy setup for reference using probe laser, SAS-C: saturated absorption spectroscopy setup for control laser locking.

The experimental arrangement for VS-BPS is different than that of a BPS setup. A separate strong control laser is used in VS-BPS technique for pump beams, as compared to the same laser used in BPS for pump and probe beams. The probe beams (Pr1 and Pr2) after passing through Glan-Thompson (GT) prism (used as an output analyzer) are detected by a photodetector pair (PDP) placed in electronic difference configuration. The intensities of probe beams Pr1 and Pr2 have small difference (~4%) due to their generation from different surfaces of the glass plate used in the experiment. The individual PS signals are made equal by controlling the intensities falling on the individual photodetectors. This helps in minimizing the Doppler background in the signal. In the VS-BPS technique, each pair of counter-propagating pump and probe beam generates a dispersion-like signals, and two such signals subtracted using a difference amplifier (DA) give the resultant VS-BPS signal with the negligible background in the spectrum. The angle ($\theta'$) between the axis of GT and that of the polarizer (PBS3) is adjusted so that the VS-BPS signal is optimized ($\theta' = \pi/4$) to obtain sharp dispersive slope in signal for frequency stabilization [14].

The VS-BPS signal is expected to depend on number density of Kr* atoms in the cell. Hence, the parametric dependence of the signal on RF power applied to the sample may be important in case of metastable Krypton atoms.

## 4. RESULTS AND DISCUSSION

### (a) Generation of VS-BPS signal in Kr* gas:

In order to generate VS-BPS signal in the Kr* atoms, the control laser is locked to 6 MHz blue detuned to the resonance transition $4p^55s[3/2]_2$ to $4p^55p[5/2]_3$ of $^{84}$Kr* using side-locking technique with SAS signal. The frequency of the probe laser is scanned around the same transition to generate sharp dispersion like profile for the VS-BPS signal. The control laser is detuned by adjusting the DC voltage on Piezo-electric Transducer (PZT) attached to the grating of the ECDL system.

Fig. 5 shows the VS-BPS signal generated using our setup. Since, the control laser is locked at 6 MHz detuned from the resonance of $^{84}$Kr*, the VS-BPS signal is red-detuned by same amount from the resonance, which corresponds to the velocity ~ 4.9 m/s of $^{84}$Kr* atoms. In accordance with equations (9)-(11), the $^{82}$Kr* atoms with velocity $v_z$ ~ 51.9 m/s show VS-BPS at around -111 MHz and the $^{86}$Kr* atoms with velocity $v_z$ ~ -48.7 m/s show that around 100 MHz detuning (Fig. 5(a)). In the absence of control laser beam, the signals from Pr1 and Pr2 are nullified when subtracted by the difference amplifier as shown in Fig. 5(b). The SAS signal shown in Fig. 5(c) is generated using probe laser and used to observe relative frequency shift of the dispersive signal. We are mainly interested in the VS-BPS obtained near resonance in this study because of its application to the $^{84}$Kr*-MOT.

### (b) Dependence of VS-BPS signal on RF and control laser powers:

For metastable $^{84}$Kr* atoms, it is important to study the effect of applied RF power on the spectral features of VS-BPS signal. Recently, the optimization of PS has been studied for alkali (Rb) atoms [34] as well as metastable inert gas (He) atoms [35]. The study mainly shows the change in amplitude and slope of PS signal with respect to the control laser power and temperature We identify here that, the VS-BPS signal is highly sensitive to RF power applied to the metastable inert gases. Thus, we have studied the VS-BPS amplitude and sharpness of slope of the this signal as a function of RF power for a constant value of control laser power in subsequent sections.

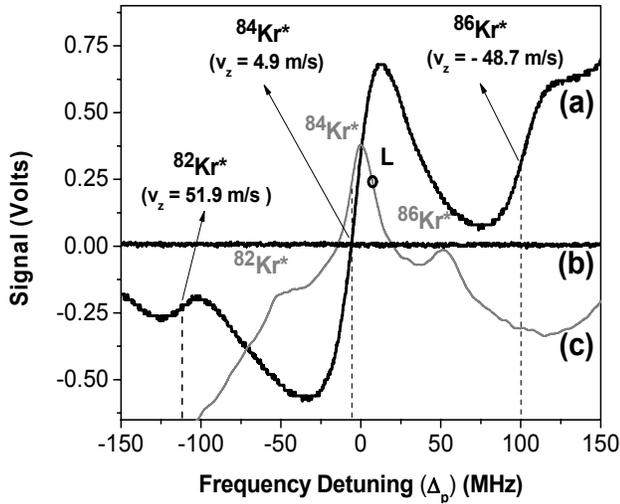

Fig.5. The observed VS-BPS for Kr* atoms when the control laser detuning is $\Delta_c = 6$ MHz, with control laser beam power 9 mW (curve (a)) and 0mW (curve (b)) and ~ 56 mW of RF power. The SAS signal shown (curve (c)) is generated from the setup SAS-R for frequency reference. The L indicates the position where control laser frequency has been locked using a SAS signal generated from the setup SAS-C (Fig. 4).

The amplitude increases faster with RF power (Fig. 6(A), (a)). It reaches at the maximum value for a certain value of RF power. A further increase in the RF power results in slow decrease in amplitude of the signal. Also the maxima in each of the curves shifts towards lower RF powers as we reduce the control laser power. The plots show that the change with control laser power in signal amplitude at lower RF power values is very small compared to that at higher RF power values. For extreme high values of RF power, this change in signal amplitude becomes negligible. At very large RF power, the signal itself is very small, for example, for control laser powers 9 mW (Fig. 6 (A), (d)) and 7 mW (Fig. 6 (A), (e)). The overall signal amplitude always decreases with decrease in control laser power as shown in Fig. 6 (A). The nature of such curves can be understood, as follows. The initial increase in RF power increases the number of Kr atoms in the metastable state. The increase in the number of metastable atoms gets saturated with further increase in the RF power due to various loss mechanisms including ionization processes.

On other hand, it is interesting to observe that, for a certain value of control laser power, the slope (V/MHz) of the VS-BPS signal (Fig. 6 (B)) first increases with RF power (curve (a)), to attain its maximum, and then decreases with further increase in RF power as the signal starts to become broadened due to higher rate of collisions. At extreme high RF power values the signal and its slope both change very little with change in RF, and overall values of amplitude and slope are very small in this regime. As the control laser power is decreased further, the maximum slope value increases and also the shifts towards lower values of RF power (Fig. 6 (B) (b), (c)). This trend is continued upto the values of control laser power 9 mW, here we get largest slope. After reaching a maximum value for the slope of the signal, a further decrease in control laser power (Fig. 6 (B) (e)) does not significantly changes the power broadened linewidth of transition linewidth although it continues to reduce the signal amplitude due to reduced pumping rate of the Kr* atoms to the excited state. This results in optimum value for the steepness of the slope with the control laser power. Therefore, it is important to optimize the experimental parameters such as RF power and control power to achieve largest slope of the locking signal.

The significant dependence of VS-BPS signal on RF and pump beam power values makes it interesting to investigate the stability of the probe laser locked at two different points of the same curve of Fig. 6. This study has been discussed in the next subsection.

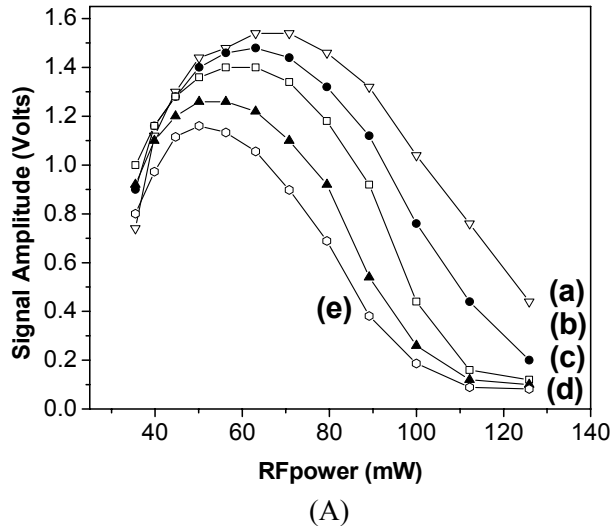

(A)

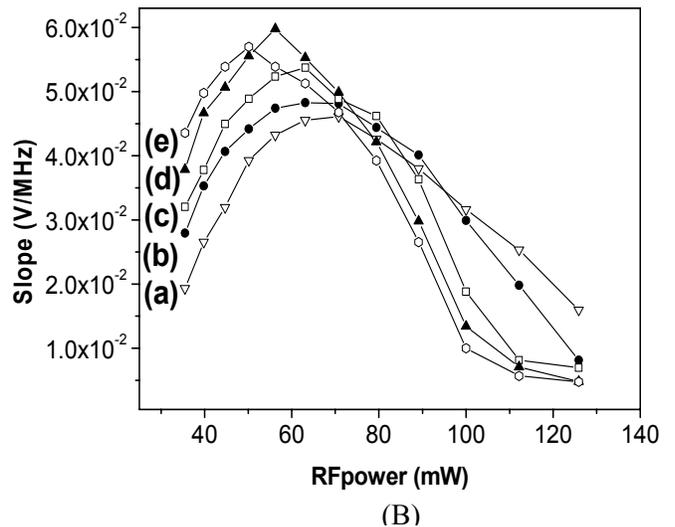

(B)

Fig. 6. The plot of (A) VS-BPS signal amplitude and (B) VS-BPS slope as a function of applied RF power, for the probe laser power value of ~ 0.5 mW.and control laser power values (a) 14 mW (b) 13 mW (c) 12 mW (d) 9 mW (e) 7 mW.

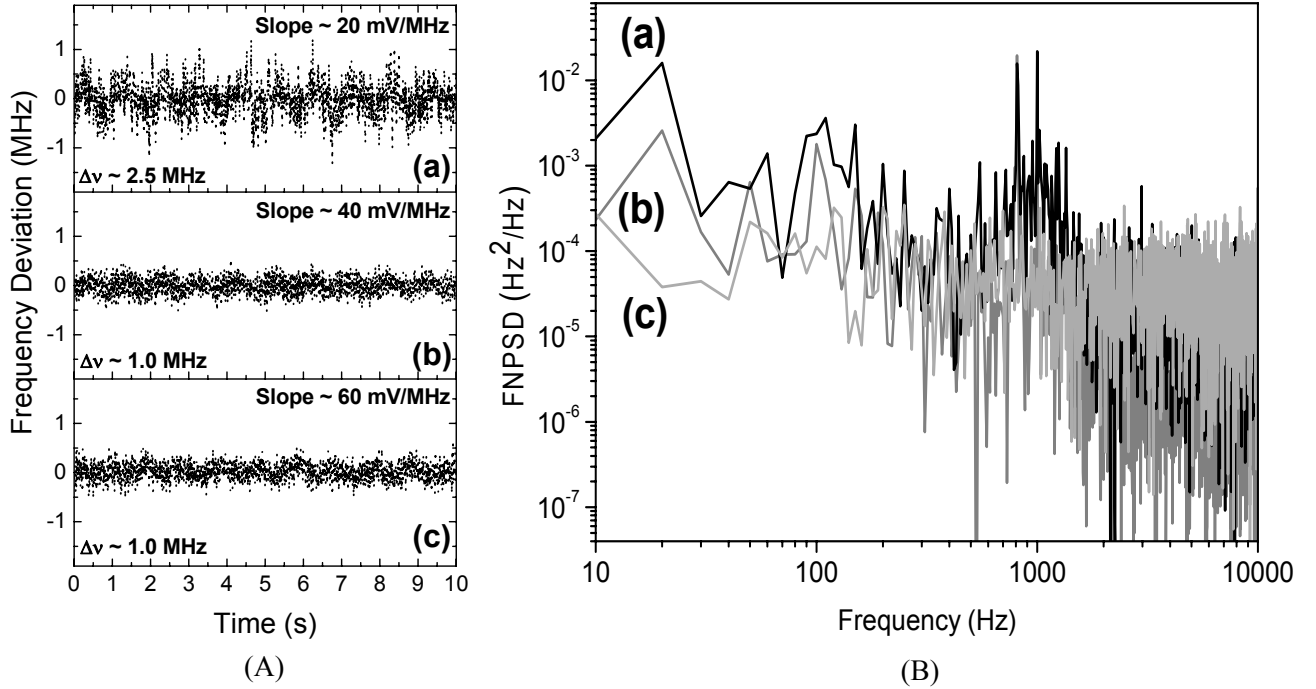

Fig.7. (A) The recorded error signals after probe laser locked with VSBPS signal for the different values of signal slope; (a): ~ 20 mV/MHz (Δν ~ 2.5 MHz), (b): ~ 40 mV/MHz (Δν ~ 1.0 MHz) and (c): ~ 60 mV/MHz (Δν ~ 1.0 MHz). (B) The frequency noise power spectral density (FNPSD) for the corresponding cases in (A).

(c) Stabilization of probe laser using VS-BPS signal:

The frequency stability of the probe laser is investigated using the VS-BPS locking signal of two different slopes while keeping the PID controller settings unchanged. It is clearly observed from Fig. 7 (A) that the frequency fluctuations shows difference in frequency deviation for locking using different slopes (i.e. for 20mW/MHz and 40 mW/MHz). We also locked the laser with further higher slope of 60 mW/MHz and observed no further appreciable reduction in the frequency deviation. This is perhaps due to fixed PID setting which we mentianed during the measurements. The frequency domain analysis of the error signals is also performed by calculating the frequency noise power spectral density (FNPSD), given by [36],

$$S_{\Delta\nu}(f) = 2\int_0^\infty \langle \Delta\nu(t+\tau)\Delta\nu(t)\rangle e^{-i2\pi f\tau} d\tau \quad (13)$$

where f is the Fourier frequency, Δν (t) is the instantaneous frequency fluctuations and τ is sampling time. The comparison of of both conditions shows that noise contribution using steeper VS-BPS signal is always smaller for all Fourier frequencies (Fig. 7 (B)). It is clear from these observations that the probe laser has smaller frequency fluctuations when locked using VS-BPS signal with steeper slope. Therefore it is important to choose experimental parameters such that one gets a better signal slope for frequency locking. Further, we have compared the performance of VS-BPS locking technique presented here with that of the conventional SAS locking technique. For this purpose, we locked our probe laser (i.e. cooling laser for MOT) using both these techniques, using the setup as shown schematically in Fig. 4. The VS-BPS signal is comparatively sharper than the SAS signal, particularly for a closed transition which is generally employed for atom cooling. Moreover, the cross-over peaks in a VS-BPS signal are much smaller in amplitude than in a SAS signal [24]. Hence, the resonance peaks are better resolved using VS-BPS technique. The error signals obtained after locking the probe laser with SAS (slope ~ 30 mV/MHz) and VS-BPS (slope ~ 60 mV/MHz) techniques are shown in Fig. 8. While locking the probe laser with VS-BPS, the control laser was locked with SAS signal. Nearly same RF power (~ 56 mW), control/pump laser power (~ 9 mW) and probe laser power (~ 0.7 mW) are used in the Krypton cell for SAS setup as well as for VS-BPS setup. The error signals after frequency locking were recorded for more than an hour using LABVIEW based data acquisition system. The system records the voltage change in the interval of 5 seconds. The results in Fig. 8 clearly show that the probe laser frequency fluctuations after locking with VS-BPS signal (Δν ~ 1.0 MHz) are considerably smaller than those obtained after locking with SAS signal (Δν ~ 2.0 MHz). Therefore, the VS-BPS signal gives better perfoprmance than the SAS signal. We note here that for locking probe laser with VS-BPS, the corresponding control laser is locked with SAS technique using a separate setup (SAS-C shown in Fig. 4). The smaller fluctuations observed with VS-BPS technique than those observed with SAS for locking probe laser indicates that frequency fluctuations of control laser have no observable effect on VS-BPS locked probe laser fluctuations. This is because in VS-BPS scheme, the probe laser is locked with a signal of larger slope than the slope of SAS signal for locking the control laser. An observation similar to this is reported earlier also where a probe locked with norrower EIT signal (with larger slope) showed better stability than the SAS locked pump laser [20]. We also note here that the experimental setup was not put under any enclosure to reduce the ambient temperature fluctuations in the laboratory. However, the frequency drift observed over a time period of ~ 1 hour was less than ~ 0.2 MHz when temperature in the laboratory changed within ± 0.5 °C . The small drift is mainly due to the reduction of the background level in the frequency locking signal.

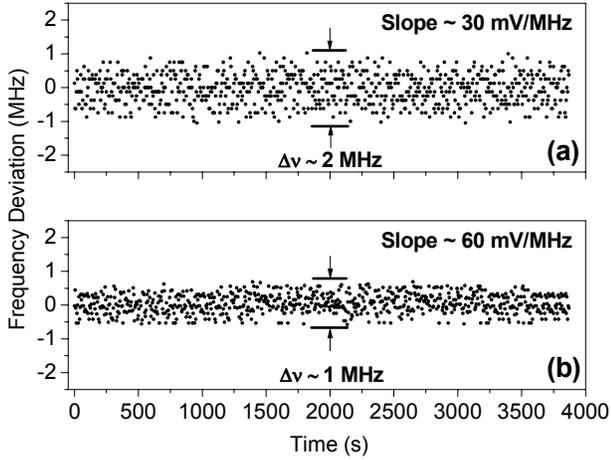

Fig. 8. The recorded error signals after locking the probe laser using different spectroscic techniques: (a) SAS and (b) VS-BPS.

    Since a wide range of velocity class of atoms is available to simultaneously interact with the control and probe lasers, we can get the VS-BPS signal over a large range of control laser frequency detuning. This provides an ability to tune the VS-BPS signal over a wide range of frequency determined by the frequency lock-point of the control laser. In order to show the tunabillity of VS-BPS signal with the control laser frequency, we recorded VS-BPS signal for different lock points of control laser frequency. The Fig. 9 illustrates the tunable nature of VS-BPS signal around the $^{84}Kr^*$ transition. For example, if the control laser for pump beams is locked at ~ 6 MHz blue detuned from the transition, the probe laser, when scanned, gives VS-BPS signal at ~ 6 MHz red detuned side (Fig. 9 (c)). Similarly, VS-BPS signal can be obtained on blue side of the transition, if control laser is locked at the red-detuned side. We observed that probe laser VS-BPS signal zero-crossing can tuned in the range of + 30 MHz to – 30 MHz with respect to resonance frequency of $^{84}Kr^*$. This is a large tunability that can be obtained along with the robust locking of an ECDL stystem.

    Thus our proposed VS-BPS technique makes use of two different aspects of frequecy locking. First is robust locking due to use of a BPS signal with large slope and second is large tunabilty obtainable due to control laser frequency dependent zero-crossing of the BPS signal. The tuning range here is also dependent on the type of technique used for locking of the control laser. Thus, we believe that using a widely tunable locking technique for control laser, such as Dichroic Atomic Vapour Laser Lock (DAVLL) [37] and Doppler-free Dichroic Lock (DFDL) [38,39], one can further increase the range of tunability for VS-BPS signal to lock the probe laser (i.e. cooling laser for MOT). The techniques that can provide further steeper slope include EIT resonances [19] and generation of odd-derivatives of the signal using lock-in amplifier [20], can also be a substitite to BPS in our proposed scheme.

    The VS-BPS signal has been used to lock the probe laser (i.e. cooling laser in a MOT setup) at the required frequency. We can tune this cooling laser in locked condition itself, to precisely control the number of cold atoms or temperature in a MOT as discussed in the nex sub-section.

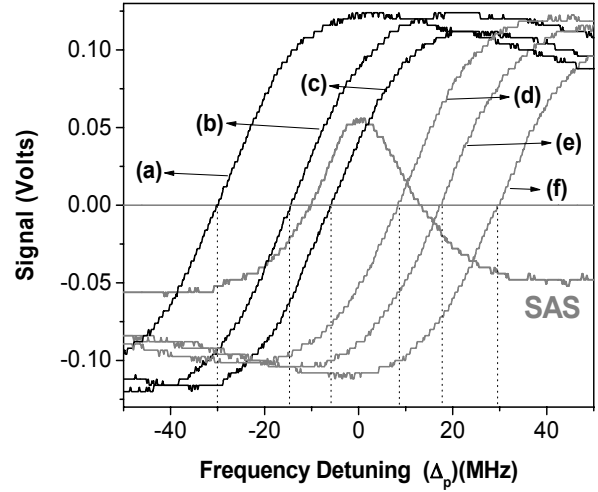

Fig. 9. The experimentally measured VS-BPS signals as function of probe laser detuning ($\Delta_p$) for different locking positions (a)-(f) of control (pump) beam frequency detuning ($\Delta_c$). The control beam frquency detuning values are (a) -30 MHz, (b) -14 MHz, (c) -6 MHz, (d) 8 MHz, (e) 17 MHz and (f) 30 MHz. The SAS (Gray) signal shown is generated from the SAS-R setup in Fig. 4.

### (d) Application of VS-BPS for Kr*-MOT setup:

As we have demonstrated in previous subsection, the VS-BPS signal can be tuned in locked condition. Hence, it can be applicable for tunable frequency stabilization of the cooling laser in our Kr*-MOT. Fig. 10 shows the schematic of the experimental setup for the laser cooling and trapping of Kr* atoms. The Kr gas enters into the inlet chamber C1 ($10^{-3}$ Torr). The Kr atoms are excited to metastable state $4p^55s[3/2]_2$ by a RF radiation (frequency ~ 30 MHz) in the glass tube with dimensions 10 mm (diameter) and 150 mm (length). The Kr* atoms are collimated through a collimation tube before entering into chambers C2 ($10^{-5}$ Torr) and C3 ($10^{-6}$ Torr). The Kr* atoms pass through Zeeman slower (the magnetic field in coil varies from 195 G to 20 G along the length of 80 cm) and enter into MOT chamber ($10^{-8}$ Torr) with speed (~ 20 m/s). The probe laser in VS-BPS setup is used as cooling for MOT by dividing the main output beam of the laser in various MOT-beams (six beam configuration with power ~ 8 mW and $1/e^2$ size ~ 5 mm). The magnetic field gradient of the pair of the coils (in anti-Helmholtz configuration) is approximately 10 G/cm. The even isotopes of Krypton, such as $^{84}Kr^*$ used here for cooling and trapping, are without hyperfine structures. Hence no repumping laser beam is required for these isotopes.

    The cooling laser beam was locked using VS-BPS signal generated by optimizing RF power and control laser power carefully. By changing the control laser frequency, the lock-point for the probe laser (i.e. cooling laser) can be tuned, enabling the variation in the detuning of cooling laser. By varying the cooling laser detuning,, the number and temperature of cold atoms in the MOT can be varied. The number of cold atoms N, in the atomic cloud was measured by imaging it on CCD camera and using the relation[40],

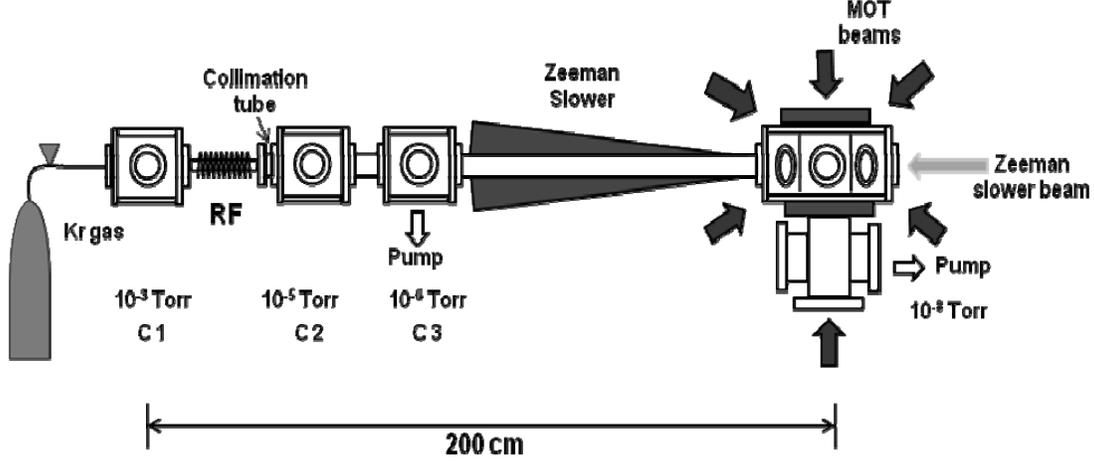

Fig. 10. Schematics of the setup for Kr* atoms laser cooling and trapping. It includes vacuum chambers C1, C2 and C3 with pressures $10^{-3}$ Torr, $10^{-5}$ Torr and $10^{-6}$ Torr. The MOT chamber with pressure $10^{-8}$ Torr is connected to other chambers through Zeeman slower.

$$N = \frac{8\pi\left(1 + 4\left(\frac{\Delta_p}{\Gamma_n}\right)^2 + 6S_0\right)}{6\Gamma_n S_0 t_{exp} \eta \Omega} N_{count} \quad (14)$$

where, $N_{count}$ is the total number of counts in the image, $t_{exp}$ is the CCD exposure time, $\eta$ is the quantum efficiency of CCD, the saturation parameter $S_0 = I_0/I_{sat}$, $I_{sat}$ is the saturation intensity of transition (~1.36 mW/cm$^2$), $I_0$ is the intensity of each of the cooling beam in the MOT and $(\Gamma_n/2\pi)=5.56$ MHz is the natural linewidth of $^{84}$Kr* transition, $\Delta_p$ is the detuning of the cooling laser beam, and $\Omega$ is a solid angle for the collection of fluorescence by the CCD. The curve (a) in Fig. 11 shows the dependence of number of cold atoms on the detuning of the cooling laser. We precisely tuned our cooling laser using VS-BPS to the intervals of 2 MHz in locked condition of the lasers. The number of cold atoms is found to attain an optimum value of ~ $5.3 \times 10^4$ for $\Delta_p$ ~ -6 MHz.

The temperature of the cloud is estimated by measuring the size of atom and the spring constant ($\kappa$) which is given by [41],

$$\kappa = \frac{8\mu A \hbar \left(\frac{2\pi}{\lambda}\right)^2 \Delta_p S_0}{\Gamma_n \left[1 + S_0 \left(\frac{2\Delta_p}{\Gamma_n}\right)^2\right]^2} \quad (15)$$

where, μ is the magnetic moment, A is the magnetic field gradient and λ the resonant wavelength. According to the equipartition theorem [41],

$$\kappa x_{rms}^2 = k_B T \quad (16)$$

where, $\kappa$ and $x_{rms}$ are the spring constant and root mean square radius of the cold atom cloud respectively. By substituting the experimentally obtained value of $x_{rms}$ for the corresponding value of $\Delta_p$ the temperature is calculated and plotted as a function of the cooling laser detuning. From curve (b) in Fig. 11, the temperature decreases from ~ 900 μK to ~ 380 μK when the detuning is increased from $\Delta_p$ ~ -4 MHz to ~ -12 MHz.

These results on the change in number and temperature of cold atoms with cooling laser frequency detuning agree well with earlier reported studies on MOT characterization [42, 43].

## 5. CONCLUSION

We have generated a velocity selective bi-polarization spectroscopy (VS-BPS) signal for locking a cooling laser for metastable Krypton atoms MOT. The parametric dependence of VS-BPS signal on different values of RF power and pump laser power (i.e. control laser) power used in the experiment has been studied. The frequency stability of a diode laser locked with VS-BPS signal has been found to be better than the frequency stability of the laser locked with a saturated absorption spectroscopy (SAS) signal. The VS-BPS signal has been finally used for tuning and stabilization of the frequency of a cooling laser used for metastable Krypton atoms MOT. The large frequency tuning range available with VS-BPS scheme makes it attractive to use in a laser cooling setup.

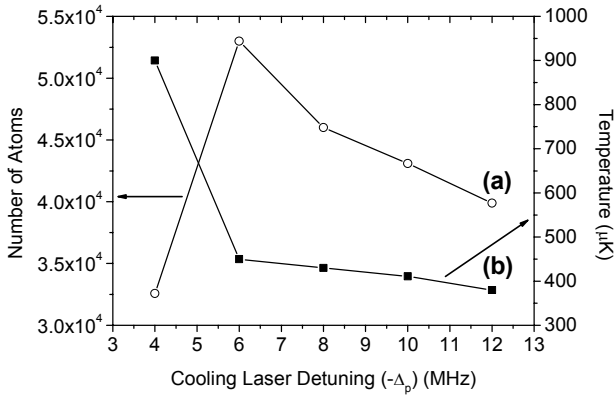

Fig. 11. The variation in (a) number of cold atoms and (b) temperature in atom cloud in Kr*-MOT with detuning of the cooling laser frequency.

ACKNOWLEDGEMENTS

We are thankful to Mandar J. Joshi, Laser Electronics Support Division, for the data acquisition system used in this paper. YBK is also thankful for Department of Atomic Energy post doctoral fellowship.